\documentclass[a4paper,10pt,twoside]{cpc-hepnp}

\pdfoutput=1
\usepackage{multicol}
\usepackage{graphicx}
\usepackage{booktabs}
\usepackage{amssymb,bm,mathrsfs,bbm,amscd}
\usepackage[tbtags]{amsmath}
\usepackage{lastpage}

\begin{document}



\title{Research on the Design and Simulation of the CSRe Stochastic Cooling System}

\author{%
      HU Xue-Jing(胡雪静)$^{1,2;1)}$\email{huxuejing@impcas.ac.cn}%
\quad YUAN You-Jin(原有进)$^{1}$
\quad WU Jun-Xia(武军霞)$^{1}$ \\
\quad ZHANG Xiao-Hu(张小虎)$^{1,2}$
\quad JIA Huan(贾欢)$^{1,2}$
}
\maketitle

\address{%
$^1$ Institute of Modern Physics,Chinese Academy of Sciences,  Lanzhou 730000, China\\
$^2$ University of Chinese Academy of Sciences, Beijing 100049, China\\
}

\begin{abstract}
Stochastic cooling, as the different cooling mechanism from electron cooling, is designed and constructed on CSRe (experimental Cooling Storage Ring) in HIRFL (Heavy Ion Research Facility in Lanzhou) at present. The beam extracted from CSRm (main Cooling Storage Ring) is injected into the CSRe, and then the secondary beam with large momentum spread and large emittance is produced at the internal target area. This paper is designed to optimize the lattice of CSRe in order to satisfy the requirement of stochastic cooling system. Particle Tracking Method is used and both transverse and longitudinal simulation is made on the basis of the designed lattice. The results indicate that stochastic cooling is suitable for beam cooling with larger momentum spread or larger emittance, and cooling rate is preliminarily confirmed.
\end{abstract}


\footnotetext[0]{\hspace*{-3mm}\raisebox{0.3ex}{$\scriptstyle\copyright$}2013
Chinese Physical Society and the Institute of High Energy Physics
of the Chinese Academy of Sciences and the Institute
of Modern Physics of the Chinese Academy of Sciences and IOP Publishing Ltd}%

\begin{multicols}{2}

\section{Introduction}
Beam cooling aims at reducing the beam size and energy spread in a storage ring without beam loss. Stochastic cooling is a feedback system. A pickup electrode upstream senses the transverse position of the particle,then this signal is amplified in a broad-band amplifier and applied on a kicker downstream which deflects the particle by an angle proportional to its position error$^{[1]}$.

CSRe is used to cooling, deceleration and high resolution target experiment. Once the pulsed beam shoots the target, secondary beam with larger size or energy spread is produced. Compared with electron cooling, stochastic cooling is the most suitable method of cooling beam like this. If only electron cooling is used as before, the cooling time is much longer than the time with stochastic pre-cooling$^{[2]}$. Therefore, electron cooling combined with stochastic pre-cooling can reduce the cooling time effectively, which provides advantageous conditions for high effective experiments.

So far, the mass measurement experiment is carrying on HIRFL-CSR in the Institute of Modern Physics, and the high requirement of the beam quality is a new challenge to the CSR system. The success of the mass measurement experiment is a milestone in the mass standardization field. Therefore, how to get the beam cooled in the shortest time is an urgent problem presently.

Building a stochastic cooling system on CSRe has significance.

\section{Lattice Design For CSRe Stochastic Cooling System}

The lattice of the CSRe storage ring$^{[3]}$ is composed of four same deflection sections. The circumference of CSRe is 128.80112 meters. The two straight sections are used to lay the electron cooling equipment and internal target equipment respectively. CSRe is symmetrical along the short axis, and the whole lattice consists of triplet type and doublet type. The dipole of CSRe uses C type, in order to make it easy to inject beam or lay electrodes. As there is no spare space for placement of stochastic cooling equipment, pickups and kickers are installed into the dipole or quadrupole$^{[4]}$.

According to stochastic cooling rate formula (1)$^{[5]}$ and the basic bandwidth formula (2) of stochastic cooling system:
\begin{eqnarray}
\label{eq1}
\frac{1}{\tau} &=& \frac{W}{N}[{2\rm g}(1-{\tilde{M}}^{-2})-{\rm g}^{2}({\rm M}+\frac{U}{{\rm Z}^{2}})].
\end{eqnarray}
\begin{eqnarray}
\label{eq2}
W &=& \frac{1}{2\times{\rm T}_{pk}\times{\eta}\times\frac{\delta\rm p}{p}}.
\end{eqnarray}
\begin{eqnarray}
\label{eq3}
\eta &=& |\frac{1}{{{\gamma}_{t}}^{2}}-\frac{1}{{\gamma}^{2}}|.
\end{eqnarray}

W is the bandwidth of the system, N is the particle number, g is the gain, M is the mixing factor from kicker to pickup and $\tilde{M}$ from pickup to kicker. U is the signal to noise ratio and Z is the charge number.

Maximum bandwidth is calculated by three $\gamma_{t}$ of different lattices,$\gamma_{t}$ =1.395, 1.748, 2.481. The flight time from pickup to kicker $T_{pk}$ is set to 0.33μs and momentum spread ${\delta \rm p}/{\rm p}$ is ±5*$10^{-3}$. The results are shown in figure 1, 2. Since the maximum magnet rigidity of CSRe is 9 Tm, the maximum energy is 708 MeV/u (${^6}C^{12+}$). From figure 2, when the transition gamma is 1.748, the bandwidth between 400MeV/u and 650MeV/u is larger than others. So the transition gamma chosen in this paper is close to 1.8. formula (3) is the ring slipping factor as the function of transition gamma.

\begin{center}
\includegraphics[width=8cm]{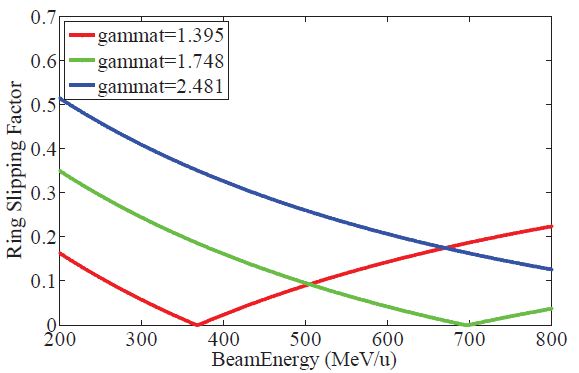}
\figcaption{\label{fig1} Ring Slipping Factor vs Transition gamma }
\end{center}
\
\begin{center}
\includegraphics[width=8cm]{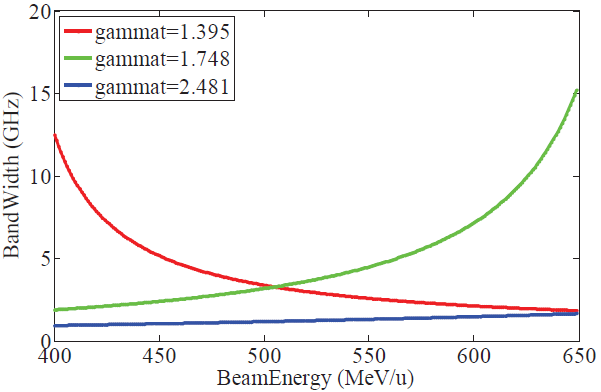}
\figcaption{\label{fig2} Maximum Bandwidth vs Transition gamma }
\end{center}
\

Stochastic cooling aims at cooling secondary particles produced at the internal target area. So the dispersion at the target is expected to be small. Figure 3 shows the twiss parameters of CSRe. Compared with the largest dispersion 15m, $D_x$ is 5.08m at the target section, which is suitable to the cooling system. Figure 4 is betatron tune distribution with different momentum spread. The red points is the distribution without using sextupole and green points with sextupole. After chromaticity correction, betatron tune area can be narrowed in a stable region. The acceptance of CSRe in the horizontal is 135 $\pi$ mm.mrad (${\delta \rm p}/{\rm p}$=0.5\%) and in the vertical direction 50 $\pi$ mm.mrad.

\
\
\begin{center}
\includegraphics[width=8cm]{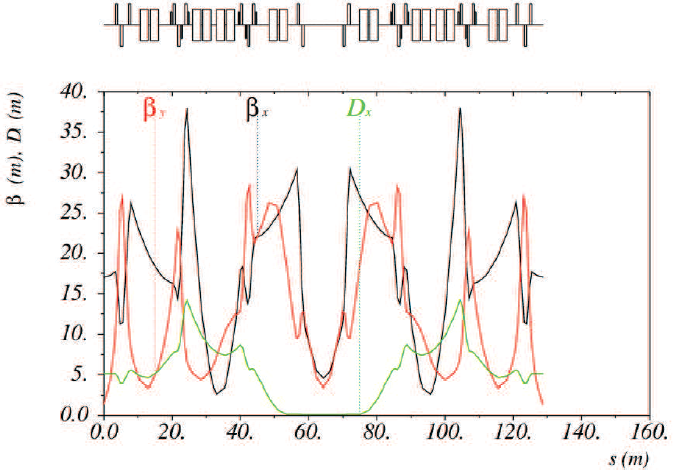}
\figcaption{\label{fig3} CSRe Twiss Parameters}
\end{center}
\
\
\begin{center}
\includegraphics[width=8cm]{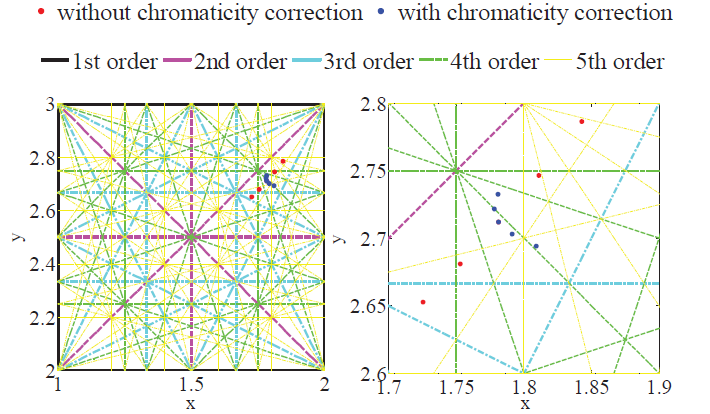}
\figcaption{\label{fig4} Betatron Tune Distribution before (red) and after (green) chromaticity correction}
\end{center}
\

\section{CSRe Stochastic Cooling Simulation and Analysis}

Since there is no spare space for the installation of pickups and kickers on CSRe, the electrodes are installed in the dipole or quadrupole. The tune shift between pickup and kicker should be (2n+1)*90°(n=0,1,…). By calculation of the designed lattice, horizontal tune shift between dipole (9) and (1) is 292.68°, and vertical tune shift between dipole (5) and (11) is 276.12°, which is suitable for transverse stochastic cooling, as show in figure 5.

Due to the transverse displacement formula of single particle$^{[6]}$:
\begin{eqnarray}
\label{eq4}
x(s) &=& \sqrt{{\bm\beta}\times\epsilon}\times\cos({\phi}+{\delta})+{\rm D}_{x}({\rm s})\times{\delta{\rm p}/{\rm p}}.
\end{eqnarray}

Horizontal displacement consists of two components. By calculation of the beam parameters at the position of dipole (5), the displacement produced by dispersion is larger than the displacement by β-function, so palmer cooling$^{[7]}$ is used at this place for longitudinal cooling. Detailed beam parameters at these positions is shown in table 1.

\end{multicols}

\begin{center}
\tabcaption{ \label{tab1} Beam Parameters for Stochastic Cooling on CSRe.}
\footnotesize
\begin{tabular*}{170mm}{@{\extracolsep{\fill}}ccccccc}
\toprule    & horizontal  & vertical & longitudinal\\
\hline
   &pickup\hphantom{0000} \hphantom{0000}kicker&pickup\hphantom{0000} \hphantom{0000}kicker&pickup\hphantom{0000} \hphantom{0000}kicker\\
\hline
L/m& 75.016-77.372\hphantom{0} \hphantom{0}10.616-12.972 & 33.060-35.416\hphantom{0} \hphantom{0}90.428-92.785 & 33.060-35.416\hphantom{0} \hphantom{0}90.428-92.785\\
$\beta_x$/m & 25.120\hphantom{0000} \hphantom{0000}19.968 & 3.438\hphantom{0000} \hphantom{00000}4.359 & 3.438\hphantom{0000} \hphantom{00000}4.359\\
$\beta_y$/m & 24.908\hphantom{0000} \hphantom{0000}3.411 & 10.075\hphantom{0000} \hphantom{0000}10.715 & 10.075\hphantom{0000} \hphantom{0000}10.715\\
$D_x$/m & 0.563\hphantom{0000} \hphantom{0000}4.624 & 7.394\hphantom{0000} \hphantom{0000}7.419 & 7.394\hphantom{0000} \hphantom{0000}7.419 \\
$A_x$/m & 127.7\hphantom{0000} \hphantom{0000}116.1 & 39.9\hphantom{0000} \hphantom{0000}79.8 & 39.9\hphantom{0000} \hphantom{0000}79.8 \\
$A_y$/m & 58.8\hphantom{0000} \hphantom{0000}30.2 & 36.6\hphantom{0000} \hphantom{0000}46.9 & 36.6\hphantom{0000} \hphantom{0000}46.9 \\
$L_{pk}$/m & 64.401 & 57.369 & 57.369 \\
$\theta$/° & 292.68(112.68) & 276.12(96.12) & \\
\bottomrule
\end{tabular*}%
\end{center}

\begin{multicols}{2}

\end{multicols}

\begin{multicols}{2}

\begin{center}
\includegraphics[width=8cm]{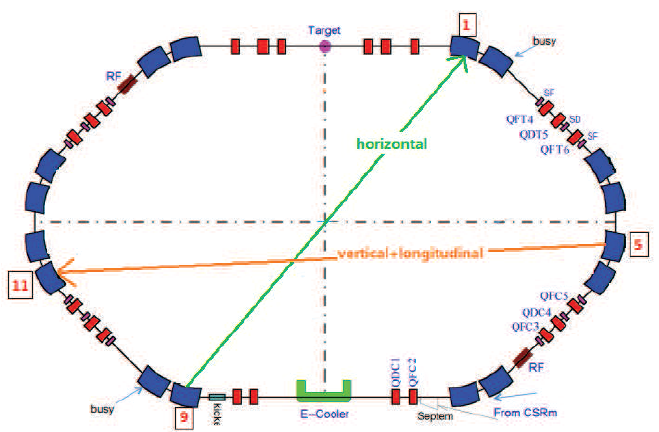}
\figcaption{\label{fig5} Layout of CSRe Stochastic Cooling System.}
\end{center}

\

\

The parameters used in the simulation is listed in table 2. Different parameters is used for calculation in order to verify the basic cooling property. Particle tracking method is used in the simulation.

\

\begin{center}
\tabcaption{ \label{tab2}Parameters of CSRe Stochastic Cooling Simulation.}
\footnotesize
\begin{tabular*}{80mm}{c@{\extracolsep{\fill}}ccc}
\toprule Beam energy & 500 MeV/u,${^{12}}C^{6+}$ \\
\hline
Particle number & 10000/3000 \\
\hline
Sample number & 383 \\
\hline
Number per sample & 26 \\
\hline
Initial horizontal emittance & 135/120/100 $\pi$ mm.mrad\\
\hline
Initial vertical emittance & 50/40/20 $\pi$ mm.mrad\\
\hline
Initial momentum spread & ±0.005/±0.002\\
\hline
Momentum cooling method & Palmer Method \\
\hline
Frequency bandwidth & 338MHz/1GHz \\
\hline
CSRe circumference & 128.801120 \\
\hline
$\gamma_t$ & 1.867281 \\
\hline
$N_{pickup}$/$N_{kicker}$ & 3/3 \\
\bottomrule
\end{tabular*}
\end{center}

\

\

Firstly, initial horizontal emittance is 135 $\pi$ mm.mrad, vertical 50 $\pi$ mm.mrad, and momentum spread is ±0.005. Figure 6, 7 and 8 shows the results under the initial conditions.

As shown in figure 6, 7, the green, red and black points corresponds to particle position in transverse phase space after 1 turn, 1000 turns and 10000 turns respectively. The momentum spread distribution during 10000 turns is shown in figure 8 and final momentum spread is about ±0.04\%. As it is clearly seen in these three pictures, particle emittance and momentum spread can be effectively reduced within 10000 turns by use of stochastic cooling$^{[8]}$.
\

\
\begin{center}
\includegraphics[width=8cm]{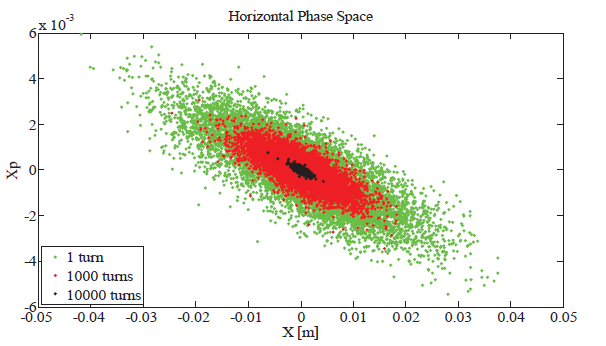}
\figcaption{\label{fig6} horizontal phase space evolution during 10000 turns ($\epsilon_{horizontal}$:135 $\pi$ mm.mrad, $\epsilon_{vertical}$:50 $\pi$ mm.mrad, ${\delta \rm p}/{\rm p}$:±0.5\%).}
\end{center}

\
\begin{center}
\includegraphics[width=8cm]{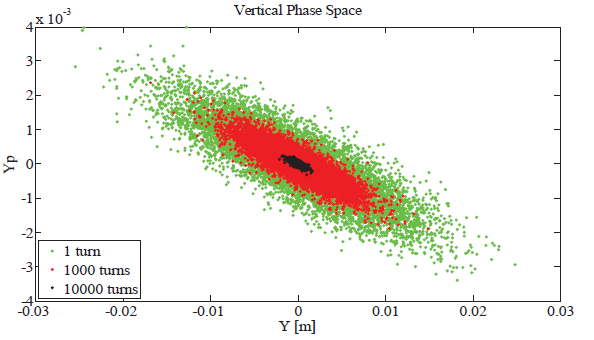}
\figcaption{\label{fig7} vertical phase space evolution during 10000 turns ($\epsilon_{horizontal}$:135 $\pi$ mm.mrad, $\epsilon_{vertical}$:50 $\pi$ mm.mrad, ${\delta \rm p}/{\rm p}$:±0.5\%).}
\end{center}

\begin{center}
\includegraphics[width=8cm]{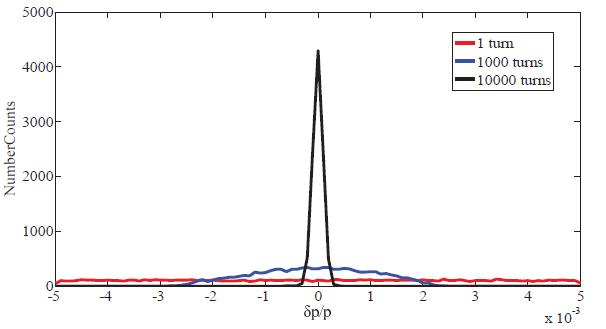}
\figcaption{\label{fig8} longitudinal phase space evolution during 10000 turns ($\epsilon_{horizontal}$:135 $\pi$ mm.mrad, $\epsilon_{vertical}$:50 $\pi$ mm.mrad, ${\delta \rm p}/{\rm p}$:±0.5\%).}
\end{center}

In order to obtain the stochastic cooling effect after 10000 turns, a comparison result between with and without cooling is shown in figure 9. The green line represents the emittance change without stochastic cooling and the blue line with cooling. It is clearly indicated that stochastic cooling is a powerful cooling method by reducing particle size effectively within a short time.

\begin{center}
\includegraphics[width=8cm]{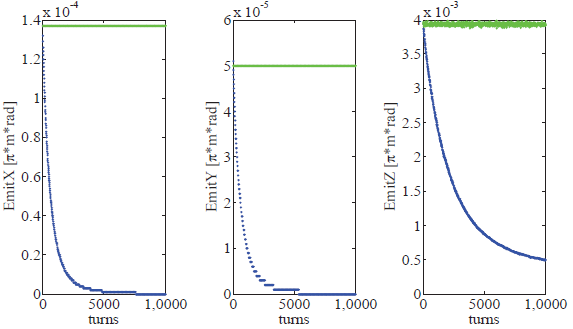}
\figcaption{\label{fig9} phase space area change during 10000 turns ($\epsilon_{horizontal}$:135 $\pi$ mm.mrad, $\epsilon_{vertical}$:50 $\pi$ mm.mrad, ${\delta \rm p}/{\rm p}$:±0.5\%).}
\end{center}

\begin{center}
\tabcaption{ \label{tab3}Stochastic Cooling Simulation Results after 2000 Turns.}
\footnotesize
\begin{tabular*}{80mm}{c@{\extracolsep{\fill}}ccc}
\toprule horizontal emittance/$\pi$ mm.mrad & 120 & 120 & 100  \\
vertical emittance/$\pi$ mm.mrad & 40 & 40 & 20 \\
momentum spread & ±0.005 & ±0.002 & ±0.005 \\
\hline
$\epsilon_x$/$\epsilon_{x0}$  & 7.38\%  & 23.77\% & 7.84\%\\
$\epsilon_y$/$\epsilon_{y0}$ & 7.32\% & 20\%  & 10\%\\
$\epsilon_z$/$\epsilon_{z0}$ & 47.47\% & 57.68\% & 47.25\% \\
Particle loss rate & 0.0\%  & 0.0\% & 0.0\%\\
\bottomrule
\end{tabular*}
\end{center}
\

The stochastic cooling simulation results after 1000 turns under other initial conditions are listed in table 3. The final results indicated that stochastic cooling is suitable for cooling particles with larger initial emittance or momentum spread.

As it is clearly shown in table 2, bandwidth in the simulation is 338MHz, and initial particle number is 10000. Then changing bandwidth from 338MHz to 1GHz and particle number from 10000 to 3000, the results are presented as follow. 

Table 4 shows the comparisons under these three conditions. It is clearly indicated that the less the particle number and the wider the bandwidth, then the higher the stochastic cooling rate, and the results is coincide with the basic stochastic cooling rate in formula (1).
\
\begin{center}
\tabcaption{ \label{tab4}Results of Stochastic Cooling Rate after 1000 turns}
\footnotesize
\begin{tabular*}{80mm}{c@{\extracolsep{\fill}}ccc}
\toprule ParticleNumber & 10000 & 10000 & 3000 \\
BandWidth & 338MHz & 1GHz & 338MHz \\
\hline
$\epsilon_x$/$\epsilon_{x0}$ & 22.63\% & 0.74\% & 0.74\% \\
$\epsilon_y$/$\epsilon_{y0}$ & 23.53\% & 2.0\% & 0.0\% \\
$\epsilon_z$/$\epsilon_{z0}$ & 66.44\% & 38.28\% & 11.05\% \\
Particle loss rate & 0.06\% & 0.08\% & 0.01\% \\
\bottomrule
\end{tabular*}
\end{center}
\

The aperture of the bending vacuum tube is 236 mm * 70 mm, and the electrodes are installed inside the beam tube in the upper, down, left, right direction respectively. The results obtained above is under the condition that the power is unlimited, and the power is about 2500 KW in the horizontal, 120 KW in the vertical and 150 MW in the longitudinal. In reality, the power cannot be so large. If the power is limited to 1 KW for three dimentions, the results are shown in table 5. As it is clearly indicated that, the cooling rate is obviously reduced by power-limited electronic system.

\
\begin{center}
\tabcaption{ \label{tab4}Limited Power for cooling after 1000 turns}
\footnotesize
\begin{tabular*}{80mm}{c@{\extracolsep{\fill}}ccc}
\toprule  &$\epsilon_x$/$\epsilon_{x0}$ & $\epsilon_y$/$\epsilon_{y0}$ & $\epsilon_z$/$\epsilon_{z0}$  \\
\hline
Unlimited power & 22.63\% & 23.53\% & 66.44\% \\
Limited power & 81.50\% & 42.06\% & 99.06\% \\
\bottomrule
\end{tabular*}
\end{center}
\
\section{Conclusion}

The results of the CSRe lattice design accord with stochastic cooling requirement. Stochastic cooling simulation is fulfilled on the basis of the designed lattice.

By analysis of table 3, stochastic cooling is suitable for beam cooling with larger momentum spread or emittance. As shown in table 4, if the bandwidth is larger or the particle number become less, the cooling rate is better. But taking other factors into consideration in reality, the bandwidth could not be very large.

From the simulation results mentioned above, after nearly 10000 turns (5.65 ms), a better cooling result both in transverse and longitudinal direction is achieved. But in reality, it is impossible to get such a cooling result in such a short time. This is because in the simulation, noise factor, power requirement as well as other hardware condition and others are not considered. For example, when the power is limited to 1KW, the cooling time is lengthened. All these factors will be considered in following simulation.

Anyhow, the result of stochastic cooling simulation shows that, it is necessary to build a stochastic cooling system on CSRe.

\end{multicols}

\vspace{-1mm}
\centerline{\rule{80mm}{0.1pt}}
\vspace{2mm}

\begin{multicols}{2}

\end{multicols}

\begin{multicols}{2}

\end{multicols}

\clearpage


\begin{thebibliography}{90}
\begin{multicols}{2}
\vspace{3mm}




\bibitem{lab6} F.Sacherer. Stochastic cooling theory.CERN Internal report ISR/TH 18-11(1978).
\bibitem{lab5} F.Nolden et al. ESR stochastic precooling. Nuclear Physics A626 (1997).

\bibitem{lab1} Xia Jiawen, Yuan Youjin, et al. CSR Overall Design. 2004.

\bibitem{lab4} Wu Junxia. Research on CSR Stochastic Cooling. Institute of Modern Physics Chinese Academy of Sciences Thesis for Doctor Degree, 2005(in Chinese).

\bibitem{lab3} D.M$\ddot{o}$hl. Stochastic cooling. CERN Report, 1995, 95-06:587;





\bibitem{lab2} S.Y.Lee.Accelerator Physics.2nd Edition.ShangHai:fudanpress. 2006. 1 --- 575.

\bibitem{lab2} Wu Junxia, Xia Jiawen, et al. HIRFL-CSRm longitudinal stochastic cooling - palmer cooling. High Energy Physics and Nuclear Physics. 2004,Vol.28,No.4.

\bibitem{lab6} D.M$\ddot{o}$hl. Phase-space cooling techniques and their combination in LEAR,Proc.Workshop on Physics at LEAR with Low-Energy Cooled Antiprotons,Erice(Plenum Press,London,1983),p.27.



\end{multicols}
\end{thebibliography}
\end{document}